\documentclass[letterpaper, 10 pt, conference]{ieeeconf}
\usepackage[utf8]{inputenc}
\IEEEoverridecommandlockouts
\usepackage[space]{cite}

\usepackage{amsmath}
\usepackage{amsfonts}
\usepackage{amssymb}
\usepackage{amsthm}
\usepackage{mathrsfs}
\usepackage{xspace}
\usepackage{xparse}
\usepackage{mathtools}

\let\labelindent\relax
\usepackage[inline]{enumitem}

\usepackage{algorithm}
\usepackage{algpseudocode}
\usepackage{setspace}

\usepackage{tabularx}
\usepackage{booktabs}
\usepackage{multirow}

\usepackage{tikz}
\usetikzlibrary{positioning}

\usepackage{epstopdf}
\usepackage{lipsum}
\usepackage{color, soul}

\newtheorem{thm}{Theorem}

\newtheorem{prob}{Problem}

\newtheorem*{BOCHNERS}{Bochner's Theorem}

\renewcommand{\Pr}{\mathop{\rm Pr}\nolimits}
\newcommand{\overbar}[1]{\mkern 1.5mu\overline{\mkern-1.5mu#1\mkern-1.5mu}\mkern 1.5mu}

\begin{document}

\title{\LARGE \bf
Approximate Stochastic Reachability for High Dimensional Systems
}

\author{%
Adam J. Thorpe,
Vignesh Sivaramakrishnan,
Meeko M. K. Oishi%
\thanks{
  This material is based upon work supported by the National Science Foundation under NSF Grant Number
  CMMI-1254990, IIS-1528047, and CNS-1836900, 
and in part by the LDRD program at Sandia Nat'l Laboratories, a multimission laboratory managed and operated by Nat'l Technology and Engineering Solutions of Sandia, LLC., a wholly owned subsidiary of Honeywell Int'l, Inc., for the US Dep't of Energy's National Nuclear Security Administration under contract DE-NA-0003525.
  Any opinions, findings, and conclusions or recommendations expressed in this
  material are those of the authors and do not necessarily reflect the views
  of the National Science Foundation,
  US Dep't of Energy, or the United States Government.
}
}

\maketitle
\thispagestyle{empty}
\pagestyle{empty}

\begin{abstract}
  We present a method to compute the stochastic reachability safety probabilities for high-dimensional stochastic dynamical systems. Our approach takes advantage of a nonparametric learning technique known as conditional distribution embeddings to model the stochastic kernel using a data-driven approach. By embedding the dynamics and uncertainty within a reproducing kernel Hilbert space, it becomes possible to compute the safety probabilities for stochastic reachability problems as simple matrix operations and inner products. We employ a convergent approximation technique, random Fourier features, in order to alleviate the increased computational requirements for high-dimensional systems. This technique avoids the curse of dimensionality, and enables the computation of safety probabilities for high-dimensional systems without prior knowledge of the structure of the dynamics or uncertainty. We validate this approach on a double integrator system, and demonstrate its capabilities on a million-dimensional, nonlinear, non-Gaussian, repeated planar quadrotor system.
\end{abstract}


\section{Introduction}

Stochastic reachability is an established verification technique which is used to compute the likelihood that a system will reach a desired state without violating a predefined set of safety constraints.
The solutions to stochastic reachability problems are broadly framed in terms of a dynamic program \cite{summers2010verification}, which scales poorly with the system dimensionality.
Methods using
approximate dynamic programming
\cite{kariotoglou2013approximate},
particle filtering \cite{manganini2015policy, lesser2013stochastic},
and abstractions \cite{soudjani2015fau} have been posed, but are limited to systems of
moderate dimensionality.
Optimization-based solutions have garnered modest computational tractability
via chance constraints \cite{lesser2013stochastic, vinod2019piecewise}, sampling
methods \cite{sartipizadeh2018voronoi, vinod2018multiple, vinod2018stochastic}, and convex optimization with Fourier
transforms \cite{vinod2018scalable, vinod2017scalable}, but are limited to linear
dynamical systems and Gaussian or log-concave disturbances.

Recent work in reachability for non-stochastic, linear dynamical systems has accommodated systems with up to a billion dimensions \cite{bak2019numerical, bak2018hylaa, bak2017hylaa}, an unprecendented size.  However, comparably scalable solutions for stochastic systems, even with considerable structure in the dynamics and in the uncertainty, remain elusive.

\emph{We propose a model-free method for stochastic reachability analysis of high-dimensional systems using a class of machine learning techniques known as kernel methods.}
We take advantage of kernel distribution embeddings \cite{smola2007hilbert}, a nonparametric learning technique that captures the features of arbitrary statistical distributions in a data-driven fashion. Distribution embeddings are amenable to generic Markov control processes, and enable efficient computation of expectations by approximating integrals via inner products.
These techniques scale exponentially with the number of samples, meaning that
they can suffer from computational complexity and memory storage requirements.
This can be prohibitive for high-dimensional systems, which may require a large number of samples in order to effectively capture the system dynamics.
The utility of distribution embeddings for the terminal-hitting time problem has been demonstrated for systems of up to 10,000 dimensions \cite{thorpe2019model}, but the jump to a million presents significant computational challenges.
%
To facilitate stocahstic reachability calculations for extremely high-dimensional systems,
we couple distribution embeddings with a technique known as random Fourier features (RFF) \cite{rahimi2008random, rahimi2009weighted}, that uses an empirical Fourier approximation to deal with high-dimensional systems.
\emph{The main contribution of this paper is an application of random Fourier features to compute an efficient model-free approximation of the safety probabilities for high-dimensional systems.}

The paper is outlined as follows. Section \ref{section: problem formulation} formulates the problem.
Section \ref{section: rkhs embeddings of distributions} outlines the theory of conditional distribution embeddings.
%
Section \ref{section: stochastic reachability using kernel distribution embeddings}
applies random Fourier features to the computation of
safety probabilities.  
In section \ref{section: numerical results}, we demonstrate our approach on two examples: a stochastic chain of integrators for validation,
and a million-dimensional, non-Gaussian, repeated planar quadrotor. 


\section{Problem Formulation}
\label{section: problem formulation}

The following notation is used throughout the paper.
For any nonempty space $\Omega$,
the indicator
$\boldsymbol{1}_{\mathcal{A}} : \Omega \rightarrow \lbrace 0, 1 \rbrace$
of $\mathcal{A} \subseteq \Omega$ is a function defined such that
$\boldsymbol{1}_{\mathcal{A}}(\omega) = 1$ if $\omega \in \mathcal{A}$, and
$\boldsymbol{1}_{\mathcal{A}}(\omega) = 0$ if $\omega \notin \mathcal{A}$.
Let $(\Omega, \mathcal{F}(\Omega), \Pr)$ define a probability space, where
$\mathcal{F}(\Omega)$ denotes the $\sigma$-algebra relative to $\Omega$, and $\Pr$ is the assigned probability measure.
When $\Omega \equiv \Re$, let $\mathscr{B}(\Omega)$ denote the Borel $\sigma$-algebra associated with $\Omega$.
Given $i \in \mathbb{N}$ random variables
$\boldsymbol{x}_{i}$,
which are measurable functions
on $(\Omega, \mathcal{F}(\Omega), \Pr)$,
let $\boldsymbol{x} = [\boldsymbol{x}_{1}, \ldots, \boldsymbol{x}_{n}]^{\top}$
be a random vector defined on the induced probability space
$(\Omega^{n}, \mathcal{F}(\Omega^{n}), \Pr_{\boldsymbol{x}})$,
where $\Pr_{\boldsymbol{x}}$ is the induced probability measure.
A stochastic process is defined as a sequence of random vectors
$\lbrace \boldsymbol{x}_{k} \rbrace_{k=0}^{N}$, $N \in \mathbb{N}$.
For a real, measurable function $\boldsymbol{x}$ on $(\Omega, \mathcal{F}(\Omega), \Pr)$,
the Lebesgue integral $\int_{\Omega} \boldsymbol{x} \Pr$
is denoted by the expectation operator
$\mathbb{E}_{\boldsymbol{x} \sim \Pr}[\boldsymbol{x}]$.


\subsection{System Model}
\label{section: system model}

We consider a Markov control process $\mathcal{H}$ \cite{summers2010verification},
which is defined as a 3-tuple:
$\mathcal{H} = (\mathcal{X}, \mathcal{U}, Q)$,
where
$\mathcal{X} \subseteq \Re^{n}$ and $\mathcal{U} \subseteq \Re^{m}$ are Borel spaces representing the state and control spaces,
and
$Q : \mathscr{B}(\mathcal{X}) \times{} \mathcal{X} \times{} \mathcal{U} \rightarrow [0, 1]$ is a stochastic kernel,
which is a Borel-measurable function
that maps a probability measure $Q(\,\cdot\, | \,x, u)$
to each $x \in \mathcal{X}$ and $u \in \mathcal{U}$
in $(\mathcal{X}, \mathscr{B}(\mathcal{X}))$.
The system evolves over a finite time horizon $k \in [0, N]$,
where the inputs are chosen from a Markov control policy
$\pi = \lbrace \pi_{0}, \pi_{1}, \ldots \rbrace$
\cite{bertsekas1978stochastic},
which is a sequence of universally-measurable maps
$\pi_{k} : \mathcal{X} \rightarrow \mathcal{U}$.

We consider the case where the stochastic kernel is unknown, but observations of the system are available.
Consider a sample $\mathcal{S} = \lbrace (\bar{x}_{i}, \bar{u}_{i}, \bar{y}_{i}) \rbrace_{i=1}^{M}$ of size $M$ drawn i.i.d. from $Q$,
such that $\bar{y}_{i} \sim Q(\cdot \,|\, \bar{x}_{i}, \bar{u}_{i})$
and $\bar{u}_{i} = \pi(\bar{x}_{i})$,
where $\pi$ is a fixed Markov control policy.
We denote sample vectors with a bar to differentiate them from time-indexed vectors.


\subsection{First-Hitting Time Problem}
\label{section: first-hitting time problem}

Let $\mathcal{K}, \mathcal{T} \in \mathscr{B}(\mathcal{X})$,
$\mathcal{T} \subseteq \mathcal{K}$,
denote the \emph{safe set} and \emph{target set}, respectively.
We define the \emph{first-hitting time safety probability} \cite{summers2010verification} as the probability that a system $\mathcal{H}$ following a control policy $\pi$ and starting at the initial condition $x_{0}$ will reach a target set $\mathcal{T}$ at some time $j \in [0, N]$ while remaining within the safe set $\mathcal{K}$ for all time $i \in [0, j - 1]$.
\begin{equation}
	r_{x_{0}}^{\pi}(\mathcal{K}, \mathcal{T}) \coloneqq
    \mathbb{E}_{x_{0}}^{\pi} \Biggl[
        \sum_{j=0}^{N} \Biggl(
            \prod_{i=0}^{j-1}
            \boldsymbol{1}_{\mathcal{K}\backslash\mathcal{T}}(x_{i})
        \Biggr)
        \boldsymbol{1}_{\mathcal{T}}(x_{j})
    \Biggr]
  \label{eqn: first-hitting probability}
\end{equation}
For a fixed Markov policy $\pi$,
we define the value functions
$V{}_{k}^{\pi} : \mathcal{X} \rightarrow [0, 1]$, $k \in [0, N]$,
via the \emph{backward recursion}:
\begin{subequations}
\label{eqn: first-hitting value functions}
\begin{align}
  V{}_{N}^{\pi}(x) &=
	\boldsymbol{1}_{\mathcal{T}}(x)
	\label{eqn: first-hitting value N} \\
  V{}_{k}^{\pi}(x) &=
	\boldsymbol{1}_{\mathcal{T}}(x) +
	\boldsymbol{1}_{\mathcal{K}\backslash\mathcal{T}}(x)
  \mathbb{E}_{\boldsymbol{y} \sim Q}[V{}_{k+1}^{\pi}(\boldsymbol{y})]
  \label{eqn: first-hitting value k}
\end{align}
\end{subequations}
Then, $V{}_{0}^{\pi}(x) = r_{x_{0}}^{\pi}(\mathcal{K}, \mathcal{T})$
for every $x_{0} \in \mathcal{X}$.
In general, computing $r_{x_{0}}^{\pi}(\mathcal{K}, \mathcal{T})$ is difficult due to the expectation in \eqref{eqn: first-hitting value k}.
We seek a representation of the stochastic kernel which enables an efficient computation of this expectation.


\subsection{Problem Statement}

We consider the following problems:

\begin{prob}
    Without direct knowledge of $Q$, use a sample $\mathcal{S}$ of observations taken from $Q$
    to compute a kernel-based approximation of \eqref{eqn: first-hitting value k} that converges in probability.
\end{prob}

\begin{prob}
    Use RFF to compute an approximation of the kernel that enables efficient computation of \eqref{eqn: first-hitting value k} for high-dimensional systems.
\end{prob}

The computational efficiencies afforded by RFF transform  \eqref{eqn: first-hitting value k} and thus \eqref{eqn: first-hitting probability} into simple matrix operations and inner products, enabling us to handle high-dimensional systems.


\section{RKHS Embeddings of Distributions}
\label{section: rkhs embeddings of distributions}

For any set $\mathcal{X}$, let $\mathscr{H}_{\mathcal{X}}$ denote a Hilbert space of real-valued functions $f : \mathcal{X} \rightarrow \Re$, with the inner product $\langle \cdot, \cdot \rangle_{\mathscr{H}_{\mathcal{X}}}$.
%
  A Hilbert space $\mathscr{H}_{\mathcal{X}}$
  is a reproducing kernel Hilbert space (RKHS)
  if there exists a positive definite
  \cite{berlinet2004reproducing} 
  kernel function
  $k_{\mathcal{X}} : \mathcal{X} \times \mathcal{X} \rightarrow \Re$
  that satisfies the following properties~\cite{aronszajn1950theory}:
    1)
  	For any $x, x' \in \mathcal{X}$,
    $k_{\mathcal{X}}(x, \cdot\,) : x' \rightarrow k_{\mathcal{X}}(x, x')$
    is an element of $\mathscr{H}_{\mathcal{X}}$.
    2)
    An element $k_{\mathcal{X}}(x, x')$ of $\mathscr{H}_{\mathcal{X}}$
    satisfies the reproducing property
    such that $\forall f \in \mathscr{H}_{\mathcal{X}}$
    and $x \in \mathcal{X}$,
    \begin{subequations}
    \begin{align}
        f(x) &=
        \langle
        k_{\mathcal{X}}(x, \cdot), f
        \rangle_{\mathscr{H}_{\mathcal{X}}}
        \label{eqn: reproducing property} \\
        k_{\mathcal{X}}(x, x') &=
        \langle
        k_{\mathcal{X}}(x, \cdot),
        k_{\mathcal{X}}(x', \cdot)
        \rangle_{\mathscr{H}_{\mathcal{X}}}
    \end{align}
    \end{subequations}
%
We define the positive definite kernels
$k_{\mathcal{X}} : \mathcal{X} \times \mathcal{X} \rightarrow \Re$ and $k_{\mathcal{U}} : \mathcal{U} \times \mathcal{U} \rightarrow \Re$,
and let $\mathscr{H}_{\mathcal{X}}$ and $\mathscr{H}_{\mathcal{U}}$
denote the RKHS
induced by $k_{\mathcal{X}}$ and $k_{\mathcal{U}}$, respectively.
Further, we define
$k_{\mathcal{X} \times \mathcal{U}} : (\mathcal{X} \times \mathcal{U}) \times (\mathcal{X} \times \mathcal{U}) \rightarrow \Re$,
via the tensor product $k_{\mathcal{X} \times \mathcal{U}}((x, u), (x', u')) =
k_{\mathcal{X}}(x, x')
k_{\mathcal{U}}(u, u')$.
Let $\mathscr{H}_{\mathcal{X} \times \mathcal{U}}$ be the associated RKHS.
%
We can also view an element $k_{\mathcal{X}}(x, \cdot) \in \mathscr{H}_{\mathcal{X}}$
as a \emph{feature map}
$\phi : \mathcal{X} \rightarrow \mathscr{H}_{\mathcal{X}}$,
such that
	$k_{\mathcal{X}}(x, x') =
	\langle
		\phi(x),
		\phi(x')
	\rangle_{\mathscr{H}_{\mathcal{X}}}$.
%
Intuitively, the feature map can be viewed as a basis function, such that a function $f \in \mathscr{H}_{\mathcal{X}}$ can be represented as a weighted sum $f(x) = \langle w, \phi(x) \rangle$ for some
possibly infinite-dimensional weight vector $w$.
However, constructing $\phi$
and computing the inner product 
explicitly can be computationally expensive or even impossible,
depending on the choice of kernel.
Instead, the inner product can be computed using $k_{\mathcal{X}}(x, x')$
directly.
This is known as the \emph{kernel trick} \cite{berlinet2004reproducing}.



\subsection{Conditional Distribution Embeddings}

For any measurable space $\mathcal{X}$,
let $\mathscr{P}$ denote the set of probability distributions on $\mathcal{X}$.
For any distribution $\mathbb{P} \in \mathscr{P}$,
if the sufficient condition
$\mathbb{E}_{\boldsymbol{x}\sim \mathbb{P}}[k_{\mathcal{X}}(\boldsymbol{x}, \boldsymbol{x})] < \infty$
is satisfied \cite{smola2007hilbert},
there exists an element $m_{\mathbb{P}}$ in the RKHS $\mathscr{H}_{\mathcal{X}}$
called the \emph{kernel distribution embedding},
\begin{align}
    \begin{split}
  \label{eqn: kernel distribution embedding}
    m : \mathscr{P} &\rightarrow \mathscr{H}_{\mathcal{X}} \\
    \mathbb{P} &\mapsto m_{\mathbb{P}} \coloneqq
    \int_{\mathcal{X}}
    k_{\mathcal{X}}(y, \cdot)
    \mathbb{P}(y)
    \mathrm{d} y
    \end{split}
\end{align}
%
This representation has several advantages.
First, if the kernel function is \emph{universal} \cite{smola2007hilbert}, the mapping is injective, meaning there is a unique element in the RKHS $\mathscr{H}_{\mathcal{X}}$
for any $\mathbb{P}, \mathbb{Q} \in \mathscr{P}$, such that
$\lVert m_{\mathbb{P}} - m_{\mathbb{Q}} \rVert_{\mathscr{H}_{\mathcal{X}}} = 0$ if and only if $\mathbb{P} = \mathbb{Q}$.
A popular kernel which satisfies these properties is the Gaussian kernel,
$k_{\mathcal{X}}(x, x') = \exp(-\lVert x - x' \rVert_{2}^{2}/2\sigma^{2})$, $\sigma > 0$.
%
Second, using the reproducing property \eqref{eqn: reproducing property},
we can compute the expectation of a function with respect to the distribution $\mathbb{P}$ as an inner product with the embedding.
\begin{align}
  \langle m_{\mathbb{P}}, f \rangle_{\mathscr{H}_{\mathcal{X}}}
  =
  \int_{\mathcal{X}}
  f(y)
  \mathbb{P}(y)
  \mathrm{d} y
\end{align}
Lastly, because $\mathbb{P}$ is typically unknown, we can compute an efficient estimate of $m_{\mathbb{P}}$. As shown in \cite{grunewalder2012conditional}, the estimate for a conditional distribution embedding is the closed-form solution of a regularized least-squares problem.

We consider mapping the stochastic kernel $Q$ into the RKHS $\mathscr{H}_{\mathcal{X}}$ \eqref{eqn: kernel distribution embedding} \cite{grunewalder2012modelling}.
Its representation in $\mathscr{H}_{\mathcal{X}}$ is given by
\begin{align}
  m_{\boldsymbol{y}|x, u} \coloneqq
  \int_{\mathcal{X}}
  k_{\mathcal{X}}(y, \cdot)
  Q(y \,|\, x, u)
  \mathrm{d} y
\end{align}
Because $Q$ is unknown, we do not have access to $m_{\boldsymbol{y}|x, u}$ directly.
Instead, we use a sample $\mathcal{S}$ drawn i.i.d. from $Q$ to compute an estimate $\hat{m}_{\boldsymbol{y}|x, u} \in \mathscr{H}_{\mathcal{X}}$
which can be found by minimizing the following optimization problem:
\begin{align}
  \label{eqn: regularized least-squares}
  \sum_{i=1}^{M}
  \lVert
    k_{\mathcal{X}}(\bar{y}_{i}, \cdot) - \hat{m}_{\boldsymbol{y}|\bar{x}_{i}, \bar{u}_{i}}
  \rVert_{\mathscr{H}_{\mathcal{X}}}^{2} +
  \lambda
  \lVert
    \hat{m}_{\boldsymbol{y}|\boldsymbol{x}, \boldsymbol{u}}
  \rVert_{\Gamma}^{2}
\end{align}
where $\Gamma$ is a vector-valued RKHS \cite{micchelli2005learning} and $\lambda > 0$ is the regularization parameter.
According to \cite{grunewalder2012conditional},
the solution to \eqref{eqn: regularized least-squares} is unique
and has the following form:
\begin{equation}
    \label{eqn: estimate}
    \hat{m}_{\boldsymbol{y}|x, u} =
    \Phi (\Psi \Psi^{\top} + \lambda M I)^{-1} \Psi k_{\mathcal{X} \times \mathcal{U}}((x, u), \cdot)
\end{equation}
The vectors $\Phi$ and $\Psi$ are known as \emph{feature vectors}, given by
\begin{align}
    \Phi &= [k_{\mathcal{X}}(\bar{y}_{1}, \cdot), \ldots, k_{\mathcal{X}}(\bar{y}_{M}, \cdot)]^{\top} \\
    \Psi &= [k_{\mathcal{X} \times \mathcal{U}}((\bar{x}_{1}, \bar{u}_{1}), \cdot), \ldots, k_{\mathcal{X} \times \mathcal{U}}((\bar{x}_{M}, \bar{u}_{M}), \cdot)]^{\top}
\end{align}
Using the estimate $\hat{m}_{\boldsymbol{y}|x, u}$, we can approximate the expectation $\mathbb{E}_{\boldsymbol{y}\sim Q}[f(\boldsymbol{y})]$ for any $f \in \mathscr{H}_{\mathcal{X}}$ as an inner product.
\begin{align}
  \langle \hat{m}_{\boldsymbol{y}|x, u}, f \rangle_{\mathscr{H}_{\mathcal{X}}}
  \approx \mathbb{E}_{\boldsymbol{y}\sim Q}[f(\boldsymbol{y})]
\end{align}
For simplicity, we can write this as
\begin{align}
  \label{eqn: expectation as inner product}
  \langle \hat{m}_{\boldsymbol{y}|x, u}, f \rangle_{\mathscr{H}_{\mathcal{X}}} = \boldsymbol{f}^{\top} \beta(x, u)
\end{align}
where $\boldsymbol{f} = [f(\bar{y}_{1}), \ldots, f(\bar{y}_{M})]^{\top}$ and $\beta(x, u) \in \Re^{M}$ is a vector of coefficients that depends on the value of the conditioning variables $(x, u) \in \mathcal{X} \times \mathcal{U}$.
\begin{align}
  \label{eqn: beta}
  \beta(x, u) = (\Psi \Psi^{\top} + \lambda M I)^{-1} \Psi
  k_{\mathcal{X} \times \mathcal{U}}((x, u), \cdot)
\end{align}
\emph{This means we can approximate the expectation of the value functions $\mathbb{E}_{\boldsymbol{y} \sim Q}[V{}_{k+1}^{\pi}(\boldsymbol{y})]$ in \eqref{eqn: first-hitting value k} as an inner product with the conditional distribution embedding estimate.}

Computing the estimate typically requires us to compute and store a matrix $G = \Psi \Psi^{\top} \in \Re^{M \times M}$, which is at least $\mathcal{O}(M^{2})$.
For large sample sizes, the storage and computation of $G$ may be prohibitive.
In order to overcome this computational challenge, we compute an approximation of the kernel itself and thus obtain a low-dimensional approximation of $G$ using a technique known as random Fourier features \cite{rahimi2008random}.


\subsection{Random Fourier Features}

As shown in
\cite{rahimi2008random},
we can reduce the computational complexity of computing \eqref{eqn: expectation as inner product}
by exploiting Bochner's theorem \cite{rudin1962fourier}.
This allows us to approximate the inner product in
\eqref{eqn: expectation as inner product}
by approximating the Fourier transform of the kernel.
\begin{BOCHNERS} \cite{rudin1962fourier}
  A translation-invariant kernel
  $k_{\mathcal{X}}(x, x') = \varphi(x - x')$ on $\mathcal{X}$
  is positive definite
  if and only if $\varphi(x - x')$
  is the Fourier transform of a non-negative Borel measure $\Lambda$.
  \begin{align}
    \varphi(x - x') &=
    \int_{\mathcal{X}}
    \exp(j \omega^{\top} (x - x'))
    \Lambda(\omega)
    \mathrm{d} \omega \\
    &= \int_{\mathcal{X}}
    \cos(\omega^{\top} (x-x'))
    \Lambda(\omega)
    \mathrm{d} \omega
    \label{eqn: real-valued property}
  \end{align}
  where \eqref{eqn: real-valued property}
  follows from the real-valued property of $\varphi$.
\end{BOCHNERS}

\noindent
Following \cite{rahimi2008random},
we construct an estimate of \eqref{eqn: real-valued property}
using a sample $\Omega = \lbrace \bar{\omega}{}_{i} \rbrace_{i=1}^{D}$
of size $D$,
such that $\bar{\omega}{}_{i}$ is drawn i.i.d. from the Borel measure $\Lambda$
according to $\bar{\omega}{}_{i} \sim \Lambda(\cdot)$.
\begin{align}
    k_{\mathcal{X}}(x, x') &\approx
    \frac{1}{D}\sum_{i=1}^D \cos(\bar{\omega}{}_{i}^{\top}(x - x'))
    \label{eqn: rff integral approximation}
\end{align}
We define a \emph{random feature map} $z : \mathcal{X} \rightarrow \Re^{D}$
such that
\begin{align}
    k_{\mathcal{X}}(x, x') \approx
    \frac{1}{D}
    \sum_{i=1}^{D}
    z_{\bar{\omega}_{i}}(x)z_{\bar{\omega}_{i}}(x')
    \eqqcolon \langle z(x), z(x') \rangle
    \label{eqn: kernel approximation rff} \\
    z_{\omega}(x) = \sqrt{2} \cos(\omega^{\top} x + b)
\end{align}
where $b$ is drawn uniformly from $[0, 2\pi]$.
Let $\hat{k}_{\mathcal{X}} \approx k_{\mathcal{X}}$ denote the kernel approximation.
Using random feature maps to approximate $k_{\mathcal{X}}$ and $k_{\mathcal{U}}$,
%
we define the feature vector $Z$,
\begin{align}
   Z &= [
     z(\bar{x}_{1}) \otimes z(\bar{u}_{1}),
     \ldots,
     z(\bar{x}_{M}) \otimes z(\bar{u}_{M})
   ]^{\top}
   \label{eqn: Z vector}
\end{align}
where $\otimes$ denotes the algebraic tensor product.
Using \eqref{eqn: Z vector},
we can approximate
\eqref{eqn: expectation as inner product} as
\begin{align}
  \langle \hat{m}_{\boldsymbol{y}|x, u}, f \rangle_{\mathscr{H}_{\mathcal{X}}} \approx
  \boldsymbol{f}^{\top} \gamma(x, u)
  \label{eqn: rff inner product approximation}
\end{align}
where $\gamma(x, u) \in \Re^{M}$ is a vector of coefficients computed using the random feature vector $Z$ in \eqref{eqn: Z vector} (cf. \cite{li2019towards}).
\begin{align}
  \gamma(x, u) &=
  (ZZ^{\top} + \lambda M I)^{-1} Z (z(x) \otimes z(u))
  \label{eqn: gamma}
\end{align}

\emph{This means we can approximate
the expectation of the value function
$\mathbb{E}_{\boldsymbol{y} \sim Q}[V{}_{k+1}^{\pi}(\boldsymbol{y})]$
in \eqref{eqn: first-hitting value k}
as an inner product of random feature maps.}
Note that the matrix
$H = Z Z^{\top} \in \Re^{D \times D}$
has lower dimensionality than $G$ if $D < M$, making it more computationally efficient to compute and store.
As remarked in \cite{rahimi2008random}, evaluating a function using the kernel trick requires $\mathcal{O}(Md)$ operations, where $d$ is the dimensionality of the data, whereas RFF only requires $\mathcal{O}(D + d)$ operations.


\section{Approximate Stochastic Reachability}
\label{section: stochastic reachability using kernel distribution embeddings}

With the conditional distribution embedding $m_{\boldsymbol{y}|x, u}$, the value function
in \eqref{eqn: first-hitting value k} can be written as
\begin{align}
  V{}_{k}^{\pi}(x) =
  \boldsymbol{1}_{\mathcal{T}}(x) +
  \boldsymbol{1}_{\mathcal{K}\backslash\mathcal{T}}(x)
  \langle m_{\boldsymbol{y}|x, u}, V{}_{k+1}^{\pi} \rangle_{\mathscr{H}_\mathcal{X}}
  \label{eqn: value function as inner product}
\end{align}
With the estimate $\hat{m}_{\boldsymbol{y}|x, u}$ and using the RFF approximation in \eqref{eqn: rff inner product approximation}, we obtain the approximation
\begin{align}
  V{}_{k}^{\pi}(x) \approx
	\boldsymbol{1}_{\mathcal{T}}(x) +
	\boldsymbol{1}_{\mathcal{K}\backslash\mathcal{T}}(x)
  \langle \hat{m}_{\boldsymbol{y}|x, u}, V{}_{k+1}^{\pi} \rangle_{\mathscr{H}_\mathcal{X}}
  \label{eqn: value function as approximate inner product}
\end{align}
We define the approximate value functions
$\overbar{V}_{k}^{\pi} : \mathcal{X} \rightarrow [0, 1]$,
$k \in [0, N-1]$ via the backward recursion
\begin{align}
  \overbar{V}_{k}^{\pi} \coloneqq
	\boldsymbol{1}_{\mathcal{T}}(x) +
	\boldsymbol{1}_{\mathcal{K}\backslash\mathcal{T}}(x)
  \langle \hat{m}_{\boldsymbol{y}|x, u}, V{}_{k+1}^{\pi} \rangle_{\mathscr{H}_\mathcal{X}}
  \label{eqn: approximate value functions}
\end{align}
where $V{}_{k}^{\pi}(x) \approx \overbar{V}_{k}^{\pi}$.
Let $\overbar{V}_{N}^{\pi} = V{}_{N}^{\pi}$.
Following \cite{thorpe2019model},
we can approximate the safety probability in
\eqref{eqn: first-hitting probability}
by approximating $\overbar{V}_{k+1}^{\pi}$
and recursively substituting it
into \eqref{eqn: approximate value functions}.
This procedure is outlined in Algorithm \ref{algo: backward recursion rff}.
Using this, we obtain the approximation $r_{x_{0}}^{\pi}(\mathcal{K}, \mathcal{T}) \approx \overbar{V}_{0}^{\pi}(x)$.


\begin{algorithm}[b!]
  \caption{Backward Recursion via RFF}
	\label{algo: backward recursion rff}

	\textbf{Input}:
	sample $\mathcal{S}$,
  evaluation point $x$,
	policy $\pi$,
	horizon $N$,
  sample $\Omega = \lbrace \bar{\omega}{}_{i} \rbrace_{i=1}^{D}$ such that $\bar{\omega}_{i} \sim \Lambda(\cdot)$
	\\
	\textbf{Output}:
	value function estimate $\overbar{V}_{0}^{\pi}(x) \approx r_{x_{0}}^{\pi}(\mathcal{K}, \mathcal{T})$

  \begin{algorithmic}[1]

		\State $\overbar{V}_{N}^{\pi}(x) \gets \boldsymbol{1}_{\mathcal{T}}(x)$
		\For{$k \gets N-1$ to $0$}
			\State Compute $\gamma(x, \pi_{k}(x))$
            from \eqref{eqn: gamma}
			using $\mathcal{S}$ and $\Omega$
			\State $\mathcal{Y} \gets [\overbar{V}_{k+1}^{\pi}(\bar{y}_{1}), \ldots, \overbar{V}_{k+1}^{\pi}(\bar{y}_{M})]^{\top}$
			\State
			$
				\overbar{V}_{k}^{\pi}(x) \gets
        \boldsymbol{1}_{\mathcal{T}}(x) +
				\boldsymbol{1}_{\mathcal{K}\backslash\mathcal{T}}(x)
        \mathcal{Y}^{\top} \gamma(x, \pi_{k}(x))
			$
		\EndFor
	    \State Return $\overbar{V}_{0}^{\pi}(x)$
	\end{algorithmic}
\end{algorithm}


\subsection{Convergence}

We now seek to characterize the quality of the approximation and analyze the conditions for its convergence.
First, we analyze the convergence of
the estimate $\hat{m}_{\boldsymbol{y}|x, u}$.
As shown in \cite{song2009hilbert}, the estimate $\hat{m}_{\boldsymbol{y}|x, u}$
converges in probability to $m_{\boldsymbol{y}|x, u}$ at a rate of $\mathcal{O}_{p}(M^{-1/4})$ if the regularization parameter $\lambda$ is decreased at a rate of $\mathcal{O}(M^{-1/2})$ (cf. \cite{grunewalder2012conditional}).

\begin{thm} \cite[Theorem~6]{song2009hilbert}
  \label{thm: embedding convergence}
  Assume $k_{\mathcal{X}}$ is in the range of $\mathbb{E}_{\boldsymbol{x}}[k_{\mathcal{X}}(\boldsymbol{x}, \cdot) \otimes k_{\mathcal{X}}(\boldsymbol{x}, \cdot)]$, then
  $\hat{m}_{\boldsymbol{y}|x, u}$ converges to $m_{\boldsymbol{y}|x, u}$
  in the RKHS norm
  at a rate of $\mathcal{O}_{p}((M \lambda)^{-1/2} + \lambda^{1/2})$.
\end{thm}

\noindent
This means we have theoretical guarantees of convergence of the embedding $\lVert m_{\boldsymbol{y}|x, u} - \hat{m}_{\boldsymbol{y}|x, u} \rVert_{\mathscr{H}_{\mathcal{X}}} \rightarrow 0$ as $M \rightarrow \infty$.
Thus, for any function $f \in \mathscr{H}_{\mathcal{X}}$, using Cauchy-Schwarz,
\begin{multline}
  \vert
  \langle m_{\boldsymbol{y}|x, u}, f \rangle_{\mathscr{H}_{\mathcal{X}}} -
  \langle \hat{m}_{\boldsymbol{y}|x, u}, f \rangle_{\mathscr{H}_{\mathcal{X}}}
  \vert \\
  \leq
  \lVert f \rVert_{\mathscr{H}_{\mathcal{X}}}
  \lVert m_{\boldsymbol{y}|x, u} - \hat{m}_{\boldsymbol{y}|x, u} \rVert_{\mathscr{H}_{\mathcal{X}}}
\end{multline}
Since $\lVert m_{\boldsymbol{y}|x, u} - \hat{m}_{\boldsymbol{y}|x, u} \rVert_{\mathscr{H}_{\mathcal{X}}}$ converges in probability according to Theorem~\ref{thm: embedding convergence},
the approximation of $f$ computed using the estimate $\hat{m}_{\boldsymbol{y}|x, u}$ also converges in probability.

Next, we consider the convergence of the RFF approximation in \eqref{eqn: rff inner product approximation}.
Convergence rates and finite-sample bounds for RFF in a generalized setting have been explored in \cite{rahimi2009weighted, rudi2017generalization, li2019towards}.
We utilize the results in \cite{rudi2017generalization},
which presents bounds for RFF in the context of least-squares problems
with Tikhonov regularization.
According to \cite[Theorem~1]{rudi2017generalization},
the approximation computed via RFF in \eqref{eqn: rff inner product approximation} has an error of $\mathcal{O}_{p}(M^{-1/2})$
if we choose $D$ according to $\mathcal{O}(M^{1/2}\log M)$
and decrease $\lambda$ at a rate of $\mathcal{O}(M^{-1/2})$.

We can use this result to show that by properly choosing $D$,
the approximate value functions converge in probability
at the rate in Theorem~\ref{thm: embedding convergence}.
Following \cite{rudi2017generalization}
and under the assumptions of Theorem \ref{thm: embedding convergence},
we present the following theorem.

\begin{thm}
  The approximate value functions $\overbar{V}_{k}^{\pi}(x)$ converge in probability to $V{}_{k}^{\pi}(x)$
  at a rate of $\mathcal{O}_{p}(M^{-1/4})$ if $D$ is chosen according to $\mathcal{O}(M^{1/2}\log M)$ and $\lambda$ is decreased at a rate of $\mathcal{O}(M^{-1/2})$.
\end{thm}

\noindent
  The proof follows by combining the convergence rates from \cite[Theorem~1]{rudi2017generalization} and Theorem~\ref{thm: embedding convergence} to obtain the result.
%
Thus, if $\vert V{}_{k}^{\pi}(x) - \overbar{V}_{k}^{\pi}(x) \vert$ has a
probabilistic error bound of $\varepsilon > 0$ at every time $k < N$,
the approximation
$r_{x_{0}}^{\pi}(\mathcal{K}, \mathcal{T}) \approx \overbar{V}_{0}^{\pi}(x)$
computed using Algorithm~\ref{algo: backward recursion rff}
converges in probability
with an error of $N \varepsilon$ \cite{thorpe2019model}.

\section{Numerical Results}
\label{section: numerical results}


\begin{figure*}
  \centering
  \includegraphics[width=\textwidth,trim=0in 0.2in 0in 0in]{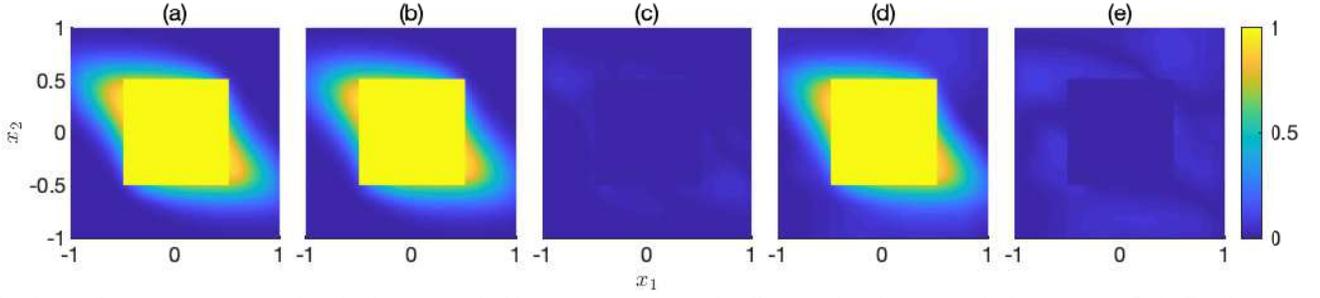}
  \caption{%
  (a) Dynamic-programming-based solution for a double integrator system with a Gaussian disturbance
  over the horizon $N = 5$.
  (b) First-hitting time safety probabilities for a double integrator system
  computed wihtout RFF
  (c) Absolute error between (a) and (b).
  (d) First-hitting time safety probabilities for a double integrator using Algorithm \ref{algo: backward recursion rff}, where $D=15{,}000$.
  (e) Absolute error between (a) and (d).}
  \label{fig: fht}
\end{figure*}

We implemented Algorithm \ref{algo: backward recursion rff} on a stochastic chain of integrators for the purposes of validation, and on a million-dimensional repeated planar quadrotor example in order to demonstrate the method for high-dimensional systems.
We generate observations
via simulation,
and then presume no knowledge of the dynamics or the structure of the uncertainty for the purposes of computing the safety probability $r_{x_{0}}^{\pi}(\mathcal{K}, \mathcal{T})$ in \eqref{eqn: first-hitting probability} using Algorithm \ref{algo: backward recursion rff}.
For all problems, we used a Gaussian kernel
$\exp(- \lVert x - x' \rVert_{2}^{2}/2 \sigma^{2})$ with $\sigma = 0.1$,
and chose $\lambda = 1$ as the default regularization parameter.
The Borel measure $\Lambda$ that corresponds to the Fourier transform of the Gaussian kernel
is a Gaussian distribution of the form
	$\Lambda(\omega) =
  \sigma^{-1}
	\exp (-\sigma^{2} \lVert \omega \rVert_{2}^{2}/2)$.

All computations were done in Matlab on a
3.8GHz Intel Xeon CPU with 32 GB RAM. Computation times were obtained using Matlab's Performance Testing Framework.
Code to generate all figures is available at
\textit{https://github.com/unm-hscl/ajthor-CDC2020}.


\subsection{Stochastic Chain of Integrators}

We consider a $2$-D stochastic chain of integrators
\cite{vinod2017scalable}, in which the input appears at the
$2^{\mathrm{nd}}$
derivative
and each element of the state vector is the discretized integral of the element that follows it. The dynamics with sampling time $T$ are given by:
\begin{align}
  \boldsymbol{x}_{k+1} =
  \begin{bmatrix}
    1 & T \\
    0 & 1
  \end{bmatrix}
  \boldsymbol{x}_{k} +
  \begin{bmatrix}
    \frac{T^{2}}{2!} \\
    T
  \end{bmatrix}
  u_{k} +
  \boldsymbol{w}_{k}
\end{align}
where $\boldsymbol{w}_{k}$ is an i.i.d. disturbance
defined on the probability space
$(\mathcal{W}, \mathscr{B}(\mathcal{W}), \Pr_{\boldsymbol{w}})$.
We consider three distributions for the disturbance:
1)
A Gaussian distribution
$\boldsymbol{w}_{k} \sim \mathcal{N}(0, \Sigma)$, where $\Sigma = 0.01 I$;
2)
A beta distribution
$\boldsymbol{w}_{k} \sim 0.1 \mathrm{Beta}(\alpha, \beta)$, with PDF
$f(x \,|\, \alpha, \beta) =
\frac{\Gamma(\alpha + \beta)}{\Gamma(\alpha) \Gamma(\beta)}
x^{\alpha-1} (1-x)^{\beta-1}$
where $\Gamma$ is the Gamma function and shape parameters $\alpha = 2$, $\beta = 0.5$; and
3)
An exponential distribution
$\boldsymbol{w}_{k} \sim 0.01 \mathrm{Exp}(\alpha)$,
with $\alpha = 3$ and PDF
$f(x \,|\, \alpha) =
\alpha \exp (-\alpha x)$.
For the purpose of validation against a known model, the control policy was chosen to be $\pi(x) = 0$
The target set and safe set are defined as $\mathcal{T} = [-0.5, 0.5]^{2}$ and
$\mathcal{K} = [-1, 1]^{2}$.

For the 2-D chain of integrators with a Gaussian disturbance,
in order to compare against a known ``truth'' model,
we computed the safety probabilities using a dynamic programming solution implemented in \cite{vinod2019sreachtools} with a time horizon of $N = 5$ (Fig. \ref{fig: fht}(a)).
Following \cite{thorpe2019model},
we then computed the safety probabilities using $\beta$ in \eqref{eqn: beta} (without RFF) using a sample $\mathcal{S}$ of size $M = 2{,}500$ (Fig. \ref{fig: fht}(b))
in order to compare against the quality of the approximation obtained using RFF.
The absolute error between the approximation and the dynamic programming solution is shown in Fig. \ref{fig: fht}(c),
and the maximum absolute error was $0.0748$.
We then generated $D=15{,}000$ frequency samples from $\Lambda(\omega)$
and computed the safety probabilities using $\gamma$ in \eqref{eqn: gamma} (with RFF) for the same sample $\mathcal{S}$ according to Algorithm \ref{algo: backward recursion rff} (Fig. \ref{fig: fht}(d)).
The absolute error between the approximation computed using RFF and the dynamic programming solution is shown in Fig. \ref{fig: fht}(e), and the maximum absolute error was $0.0907$.

\begin{figure}[t!]
  \centering
  \includegraphics[trim=0in 0.2in 0in 0in]{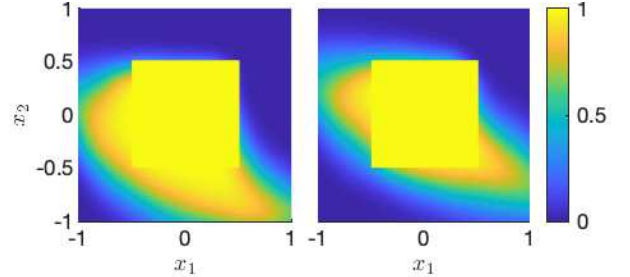}
  \caption{First-hitting time safety probabilities for a double integrator system with a beta distribution disturbance (left) and an exponential distribution disturbance (right) over the horizon $N = 50$.}
  \label{fig: fht_beta_exp}
\end{figure}

%
We then computed the safety probabilities for the same system with a beta distribution disturbance
and an exponential distribution disturbance
for a time horizon of $N = 50$.
The results are shown in Fig. \ref{fig: fht_beta_exp}.
Because Algorithm \ref{algo: backward recursion rff} is agnostic to the complexities of the disturbance, handling arbitrary disturbances is straightforward.


As expected, Algorithm \ref{algo: backward recursion rff} produced a higher error estimate of the safety probabilities due to the kernel approximation.
The quality of the approximation is dependent on $M$ and $D$, and
in some cases, the number of frequency samples $D$ required to approximate the kernel can mean Algorithm~\ref{algo: backward recursion rff} does not provide better computational efficiency.
However, when $D \ll M$, or when the system is high-dimensional, RFF can significantly reduce the computational burden.
By choosing a lower value of $D$, we exchange numerical accuracy for lower computation times of the algorithm.


\subsection{Planar Quadrotor}

We implemented Algorithm \ref{algo: backward recursion rff} on a planar quadrotor system, as well as a million-dimen\-sional repeated planar quadrotor system, comprised of 170,000 dynamically decoupled six-dimen\-sional planar quadrotors.
This problem can be interpreted as a simplification of formation control for a large swarm of quadrotors, where we compute the safety probabilities for the entire swarm as the quadrotors are controlled to reach a particular configuration.
The nonlinear dynamics of a single quadrotor are given by
\begin{align}
    \begin{aligned}[c]
        m\ddot{x} &= -(u_{1} + u_{2}) \sin(\theta) \\
        m\ddot{y} &= (u_{1} + u_{2}) \cos(\theta) - mg \\
        \mathcal{I}\ddot{\theta} &= r(u_{1} - u_{2})
    \end{aligned}
\end{align}
where $x$ is the lateral position, $y$ is the vertical position, $\theta$ is the pitch, and we have the constants
intertia $\mathcal{I} = 2$, length $r = 2$,
mass $m = 5$, and
$g = 9.8$ is the gravitational constant.


For a single quadrotor, the state space is $\mathcal{X} \subset \Re^{6}$,
with state vector given by
$\boldsymbol{z} = [x, \dot{x}, y, \dot{y}, \theta, \dot{\theta}]^{\top}$, and
the input space is $\mathcal{U} \subset \Re^{2}$,
with input vector $u = [u_{1}, u_{2}]^{\top}$.
The input is chosen to be a reference tracking controller, computed using a linearization of the system dynamics about a hover point.
%
We discretize the nonlinear dynamics in time using an Euler approximation with sampling time $T = 0.25$, and add an affine disturbance $\boldsymbol{w}$.
The disturbance
is a Markov process
with elements $\boldsymbol{w}_{k}$
defined on the probability space
$(\mathcal{W}, \mathscr{B}(\mathcal{W}), \Pr_{\boldsymbol{w}})$.
We consider two distributions for the disturbance:
1)
A Gaussian distribution
$\boldsymbol{w}_{k} \sim \mathcal{N}(0, \Sigma)$,
with variance
$\Sigma = \textnormal{diag}(1 \times 10^{-3}, 1 \times 10^{-5}, 1 \times 10^{-3}, 1 \times 10^{-5}, 1 \times 10^{-3}, 1 \times 10^{-5})$;
and
2)
A beta distribution
$\boldsymbol{w}_{k} \sim \textnormal{Beta}(\alpha, \beta)$
with shape parameters $\alpha = 2$, $\beta = 0.5$.
The beta disturbance has a non-zero mean, and can be interpreted as wind,
such that the dynamics are biased in a particular direction.
For a single planar quadrotor,
the safe set and target set are defined as $\mathcal{K} = \lbrace \boldsymbol{z} \in \Re^{6} : \vert z_{1} \vert < 1, 0 \leq z_{3} < 0.8 \rbrace$, and $\mathcal{T} = \lbrace \boldsymbol{z} \in \Re^{6} : \vert z_{1} \vert < 1, z_{3} \geq 0.8 \rbrace$.
For the repeated quadrotor system, we define the safe sets and target sets as a series of parallel tubes, such that no quadrotor may enter into the safe set of an adjacent quadrotor.
This means the quadrotors must all reach an altitude of $0.8$ while remaining within their respective tube.

\begin{figure}[t]
  \includegraphics[width=\columnwidth,trim=0in 0.2in 0in 0in]{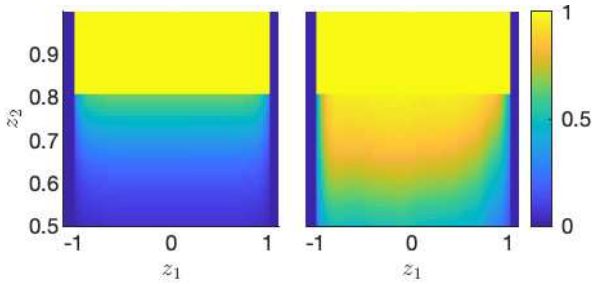}
  \caption{First-hitting time safety probabilities for a planar quadrotor system with a Gaussian disturbance (left) and with a beta distribution disturbance (right) over the horizon $N = 5$.}
  \label{fig: planar quadrotor}
\end{figure}

We first computed the safety probabilities for a single quadrotor
in order to demonstrate the capabilities of Algorithm~\ref{algo: backward recursion rff} to handle nonlinear dynamics.
We generated a sample $\mathcal{S}$ consisting of $M = 1{,}000$ observations of the single quadrotor system with a Gaussian disturbance and took $D = 15{,}000$ frequency samples from $\Lambda(\omega)$. We then computed the safety probabilities using Algorithm~\ref{algo: backward recursion rff} over a time horizon of $N = 5$ and then repeated this procedure using the beta distribution disturbance.
The results are shown in Fig. \ref{fig: planar quadrotor}.
As expected, the algorithm was able to compute the safety probabilities due to the fact that Algorithm~\ref{algo: backward recursion rff} does not exploit any knowledge of the underlying dynamics.


\begin{table}
	\caption{Computation Time}
	\label{table: computation time}
	\centering
	\begin{tabular*}{\columnwidth}{
		@{\extracolsep{\fill}} l
		@{\hspace{0.5\tabcolsep}} l
		@{\hspace{0.5\tabcolsep}} c
		@{\hspace{0.5\tabcolsep}} c
		@{\hspace{0.5\tabcolsep}} c
		}
		\toprule
    %
		  System
		& Dim. [$n$]
		& Without RFF
		& With RFF 
		& Dyn. Prog.
		\\
		\midrule
		Integrator
		& 2
		& 2.30 s
		& 22.94 s
		& 65.78 s
		\\
		Quadrotor
		& 6
		& 0.62 s
		& 15.24 s
		& --
		\\
		Quadrotor
		& 1,020,000
		& 1.23 h
		& 44.59 s
		& --
		\\
		\bottomrule
	\end{tabular*}
\end{table}


For the repeated quadrotor system, we first computed the safety probabilities without RFF in order to demonstrate the reduced computational complexity of Algorithm~\ref{algo: backward recursion rff} for high-dimensional systems.
We generated a sample $\mathcal{S}$ of $M = 1{,}000$ observations drawn i.i.d. from the stochastic kernel of the repeated quadrotor system with a beta distribution disturbance, and computed the safety probabilities (without RFF) over a time horizon of $N = 1$ from a single initial condition, to demonstrate feasibility of the approach.
We repeated this procedure 7 times and averaged the computation time over all trials to obtain an average computation time of $1.23$ hours.
We then compared this performance against Algorithm~\ref{algo: backward recursion rff} (with RFF) using the same procedure.
We generated $D = 15{,}000$ frequency samples from $\Lambda(\omega)$ and computed the safety probabilities using Algorithm~\ref{algo: backward recursion rff} over the same time horizon and the same initial condition.
%
Using the same averaging approach, we obtained an average computation time
of $44.59$ seconds.
We obtained comparable results for the Gaussian disturbance case.
As shown in Table~\ref{table: computation time}, computation time is reduced by two orders of magnitude for the high-dimensional repeated quadrotor system.


\section{Conclusions \& Future Work}

We presented an algorithm based on random Fourier features to compute the stochastic reachability
first-hitting time safety probabilities for high-dimensional Markov control processes.
This approach is applicable to arbitrary disturbances and is model-free, meaning it does not rely upon a known stochastic kernel.
We demonstrated it on a million-dimensional system to showcase
the efficiencies of the computation.
We plan to extend this to safe controller synthesis using kernel distribution embeddings.



\bibliographystyle{IEEEtran}
\bibliography{IEEEabrv, shortIEEE, bibliography.bib}

\begin{thebibliography}{10}
\providecommand{\url}[1]{#1}
\csname url@samestyle\endcsname
\providecommand{\newblock}{\relax}
\providecommand{\bibinfo}[2]{#2}
\providecommand{\BIBentrySTDinterwordspacing}{\spaceskip=0pt\relax}
\providecommand{\BIBentryALTinterwordstretchfactor}{4}
\providecommand{\BIBentryALTinterwordspacing}{\spaceskip=\fontdimen2\font plus
\BIBentryALTinterwordstretchfactor\fontdimen3\font minus
  \fontdimen4\font\relax}
\providecommand{\BIBforeignlanguage}[2]{{%
\expandafter\ifx\csname l@#1\endcsname\relax
\typeout{** WARNING: IEEEtran.bst: No hyphenation pattern has been}%
\typeout{** loaded for the language `#1'. Using the pattern for}%
\typeout{** the default language instead.}%
\else
\language=\csname l@#1\endcsname
\fi
#2}}
\providecommand{\BIBdecl}{\relax}
\BIBdecl

\bibitem{summers2010verification}
S.~Summers and J.~Lygeros, ``Verification of discrete time stochastic hybrid
  systems: A stochastic reach-avoid decision problem,'' \emph{Automatica},
  vol.~46, no.~12, pp. 1951--1961, 2010.

\bibitem{kariotoglou2013approximate}
N.~Kariotoglou, S.~Summers, T.~Summers, M.~Kamgarpour, and J.~Lygeros,
  ``Approximate dynamic programming for stochastic reachability,'' in
  \emph{Proc. European Ctrl. Conf.}, 2013, pp. 584--589.

\bibitem{manganini2015policy}
G.~Manganini, M.~Pirotta, M.~Restelli, L.~Piroddi, and M.~Prandini, ``Policy
  search for the optimal control of {Markov} decision processes: A novel
  particle-based iterative scheme,'' \emph{{IEEE} Trans. Cybern.}, vol.~46,
  no.~11, pp. 2643--2655, 2015.

\bibitem{lesser2013stochastic}
K.~Lesser, M.~Oishi, and R.~S. Erwin, ``Stochastic reachability for control of
  spacecraft relative motion,'' in \emph{Proc. IEEE Conf. Dec. \& Ctrl.}, 2013,
  pp. 4705--4712.

\bibitem{soudjani2015fau}
S.~Soudjani, C.~Gevaerts, and A.~Abate, ``{FAUST}${}^{\mathsf{2}}$: Formal
  abstractions of uncountable-state stochastic processes,'' in \emph{Int. Conf.
  on Tools and Algorithms for the Construction and Anal. of Syst.}, 2015, pp.
  272--286.

\bibitem{vinod2019piecewise}
A.~Vinod, V.~Sivaramakrishnan, and M.~Oishi, ``Piecewise-affine
  approximation-based stochastic optimal control with {Gaussian} joint chance
  constraints,'' in \emph{Proc. Amer. Ctrl. Conf.}, 2019, pp. 2942--2949.

\bibitem{sartipizadeh2018voronoi}
H.~Sartipizadeh, A.~Vinod, B.~A{\c{c}}{\i}kme{\c{s}}e, and M.~Oishi,
  ``{Voronoi} partition-based scenario reduction for fast sampling-based
  stochastic reachability computation of {LTI} systems,'' \emph{Proc. Amer.
  Ctrl. Conf.}, pp. 37--44, 2018.

\bibitem{vinod2018multiple}
A.~Vinod, B.~HomChaudhuri, C.~Hintz, A.~Parikh, S.~Buerger, M.~Oishi,
  G.~Brunson, S.~Ahmad, and R.~Fierro, ``Multiple pursuer-based intercept via
  forward stochastic reachability,'' in \emph{Proc. Amer. Ctrl. Conf.}, 2018,
  pp. 1559--1566.

\bibitem{vinod2018stochastic}
A.~Vinod, S.~Rice, Y.~Mao, M.~Oishi, and B.~A{\c{c}}{\i}kme{\c{s}}e,
  ``Stochastic motion planning using successive convexification and
  probabilistic occupancy functions,'' in \emph{Proc. IEEE Conf. Dec. \&
  Ctrl.}, 2018, pp. 4425--4432.

\bibitem{vinod2018scalable}
A.~Vinod and M.~Oishi, ``Scalable underapproximative verification of stochastic
  {LTI} systems using convexity and compactness,'' in \emph{Proc. Hybrid Syst.:
  Comput. and Ctrl.}, 2018, pp. 1--10.

\bibitem{vinod2017scalable}
------, ``Scalable underapproximation for the stochastic reach-avoid problem
  for high-dimensional {LTI} systems using {Fourier} transforms,'' \emph{IEEE
  Ctrl. Syst. Letters.}, vol.~1, no.~2, pp. 316--321, 2017.

\bibitem{bak2019numerical}
S.~Bak, H.-D. Tran, and T.~Johnson, ``Numerical verification of affine systems
  with up to a billion dimensions,'' in \emph{Proc. Hybrid Syst.: Comput. and
  Ctrl.}, 2019, pp. 23--32.

\bibitem{bak2018hylaa}
S.~Bak and P.~S. Duggirala, ``{HyLAA} 2.0: A verification tool for linear
  hybrid automaton models of cyber-physical systems,'' in \emph{{IEEE}
  Real-Time Syst. Symp.}, 2018.

\bibitem{bak2017hylaa}
------, ``{HyLAA}: A tool for computing simulation-equivalent reachability for
  linear systems,'' in \emph{Proc. Hybrid Syst.: Comput. and Ctrl.}, 2017, pp.
  173--178.

\bibitem{smola2007hilbert}
A.~Smola, A.~Gretton, L.~Song, and B.~Sch{\"o}lkopf, ``A {Hilbert} space
  embedding for distributions,'' in \emph{Int. Conf. Algorithmic Learn.
  Theory}, 2007, pp. 13--31.

\bibitem{thorpe2019model}
A.~Thorpe and M.~Oishi, ``Model-free stochastic reachability using kernel
  distribution embeddings,'' \emph{IEEE Ctrl. Syst. Letters.}, 2019.

\bibitem{rahimi2008random}
A.~Rahimi and B.~Recht, ``Random features for large-scale kernel machines,'' in
  \emph{Adv. in Neural Inf. Process. Syst.}, 2008, pp. 1177--1184.

\bibitem{rahimi2009weighted}
------, ``Weighted sums of random kitchen sinks: Replacing minimization with
  randomization in learning,'' in \emph{Adv. in Neural Inf. Process. Syst.},
  2009, pp. 1313--1320.

\bibitem{bertsekas1978stochastic}
D.~Bertsekas and S.~Shreve, \emph{Stochastic optimal control: the discrete time
  case}.\hskip 1em plus 0.5em minus 0.4em\relax Academic Press, 1978.

\bibitem{berlinet2004reproducing}
A.~Berlinet and C.~Thomas-Agnan, ``Reproducing kernel {Hilbert} spaces in
  probability and statistics,'' 2004.

\bibitem{aronszajn1950theory}
N.~Aronszajn, ``Theory of reproducing kernels,'' \emph{Trans. of the American
  Mathematical Society}, vol.~68, no.~3, pp. 337--404, 1950.

\bibitem{grunewalder2012conditional}
S.~Gr{\"u}new{\"a}lder, G.~Lever, L.~Baldassarre, S.~Patterson, A.~Gretton,
  M.~Pontil, J.~Langford, and J.~Pineau, ``Conditional mean embeddings as
  regressors,'' in \emph{Int. Conf. on Mach. Learn.}, 2012, pp. 1823--1830.

\bibitem{grunewalder2012modelling}
S.~Gr{\"u}new{\"a}lder, G.~Lever, L.~Baldassarre, M.~Pontil, and A.~Gretton,
  ``Modelling transition dynamics in {MDP}s with {RKHS} embeddings,'' in
  \emph{Int. Conf. on Mach. Learn.}, 2012, pp. 1603--1610.

\bibitem{micchelli2005learning}
C.~Micchelli and M.~Pontil, ``On learning vector-valued functions,''
  \emph{Neural Computation}, vol.~17, no.~1, pp. 177--204, 2005.

\bibitem{rudin1962fourier}
W.~Rudin, \emph{{Fourier} analysis on groups}.\hskip 1em plus 0.5em minus
  0.4em\relax Wiley, 1962, vol. 121967.

\bibitem{li2019towards}
Z.~Li, J.-F. Ton, D.~Oglic, and D.~Sejdinovic, ``Towards a unified analysis of
  random {Fourier} features,'' in \emph{Int. Conf. on Mach. Learn.}, 2019, pp.
  3905--3914.

\bibitem{song2009hilbert}
L.~Song, J.~Huang, A.~Smola, and K.~Fukumizu, ``{Hilbert} space embeddings of
  conditional distributions with applications to dynamical systems,'' in
  \emph{Int. Conf. on Mach. Learn.}, 2009, pp. 961--968.

\bibitem{rudi2017generalization}
A.~Rudi and L.~Rosasco, ``Generalization properties of learning with random
  features,'' in \emph{Adv. in Neural Inf. Process. Syst.}, 2017, pp.
  3215--3225.

\bibitem{vinod2019sreachtools}
A.~Vinod, J.~Gleason, and M.~Oishi, ``{SReachTools}: a {Matlab} stochastic
  reachability toolbox,'' in \emph{Proc. Hybrid Syst.: Comput. and Ctrl.},
  2019, pp. 33--38.

\end{thebibliography}

\end{document}